\documentclass[a4paper,11pt]{article}
\usepackage{pos}

\title{Disk-jet-wind coupling from stellar mass to supermassive black holes}

\author*[a,b]{Chris Done}

\affiliation[a]{Centre for Extragalactic Astronomy, Department of Physics, Durham University,\\ 
South Road, Durham DH1 3LE, UK}

\affiliation[b]{Kavli Institute for Physics and Mathematics of the Universe (WPI), University of Tokyo,\\ 
Kashiwa, Chiba 277-8583, Japan}

\emailAdd{chris.done@durham.ac.uk}

\abstract{Black holes are the simplest possible objects, characterised by only mass and spin. We see them via accretion, so there is one more fundamental parameter which is the mass accretion rate. Here I will review how the data from both stellar and supermassive black holes can be fit into a framework where there is a major spectral transition at $\dot{m}=L/L_{\rm Edd}\sim 0.01$ where the optically thick disc is replaced by a hot flow. This dramatic spectral change also affects the expected properties of thermal and radiatively powered winds, matching the overall properties of winds seen in new XRISM data from the stellar mass binaries, though there can also be additional UV and dust driven winds in supermassive black holes. The radio data in stellar and supermassive black holes are clear that the hot flow (not the disc) connects to the radio jet, and the radio-X-ray 'fundamental plane' can be qualitatively understood if the radio quiet AGN and stellar mass black holes have low to moderate spins, 
with the jet power set as a constant fraction of the accretion power. A small fraction of AGN (radio loud) instead 
have much higher (factor $100-1000\times$) radio-to-X-ray ratio at the same black hole mass and mass accretion rates. I speculate that these have higher jet power due to high black hole spin. I review the multiple issues still remaining in this picture, most of which are connected to the geometry and nature of the X-ray corona, and the conflicting constraints on this which come from reflection spectroscopy and polarimetry.}

\FullConference{
}

\begin{document}
\maketitle

\section{Introduction}
 
We want to understand black holes accretion flows. Firstly we want to understand how they produce the radiation we see, which depends fundamentally on the geometry, dynamics and emission processes in accretion flow. Then we also want to understand how this accretion powers the collimated jet, and the wider angle outflowing winds. These two components are the major mechanisms for AGN feedback, allowing some fraction of the accretion power to couple to the host galaxy and control its starformation history. 

Progress in much of this remains frustratingly slow, mainly due to our lack of understanding of the detailed viscosity mechanism which enables accretion. Magnetic fields must be involved,
as was recognised in the original Shakura \& Sunyaev paper 50 years ago \cite{Shakura1973} (hereafter SS73) but it is still not clear even using General Relativistic, radiative, magneto-hydrodynamic (GR RMHD) simulations whether these fields are generated by the small scale Magneto-Rotational Instability (MRI) dynamo process or whether there are also larger scale ordered fields threading the flow. 
By contrast with this lack of theoretical progress, the observations are now dramatically better than 50 years ago. Here I will review how the data can be used to build a phenomenological picture of the accretion flow, and its jets and winds, informed qualitatively by theoretical models and simulation results. 

\section{Stellar Mass black holes}

Stellar mass black holes accreting from a low mass companion star in our Galaxy form a remarkably homogeneous set - plausibly they all have similar masses due to their similar orgins in core collapse supernovae, and plausibly they all have similar spins as well for the same reason. These can sometimes be extremely well fit by the standard SS73 sum-of-(colour temperature corrected) blackbody disc models, though they normally also have a tail of emission extending to higher energies. This indicates that some small fraction of the emission is dissipated in optically thin material, spatially separated from the optically thick disc (where it would otherwise thermalise to the disc temperature). As long as this coronal emission is small, less than $\sim 10$\% of the total accretion power, then the spectra can be fit well with the SS73 disc predictions (with some corrections for Compton scattering and atomic absorption in the photosphere of the disc) \cite{Done2007}.

Not only do the individual spectra match, but their long term variability also matches. The peak disc temperature $T\propto L_{\rm disc}^{1/4}$, where $L_{\rm disc}$ is the disc luminosity, as required for a constant inner disc radius set by the innermost stable circular orbit of General Relativity. Additionally, the timescales for this mass accretion rate change also matches with models where they are produced by a disc instability triggered by Hydrogen ionisation (DIM: 
see e.g. \cite{hameury2025}). The heating wave propagation through the previously quiescent disc material triggers accretion onto the black hole, with a rise time of a few days (see D. Russell, this volume, and \cite{Dubus2001}). The copious X-ray flux produced by accretion in strong gravity irradiates the outer disc, preventing its temperature from dipping back below Hydrogen ionisation and returning to quiescence. Instead, irradiation keeps the disc in the ionised state, so it continues to accrete. This gives a (quasi) exponential decay of the mass accretion rate as the disc is depleted, until the X-rays are too dim to maintain ionisation by irradiation and the outer disc can drop back to quiescence \cite{king1998,Dubus2001}. In the DIM model, the exponential decay timescale is set by the effective viscosity of the disc, and is fairly well matched by $\alpha\sim 0.2-0.4$ for the observed 20-40 day timescales, especially when indirect irradiation via scattering in a wind from the outer disc is also included (e.g. Dubus, this volume and \cite{Dubus2019}). Winds are now seen directly via blueshifted absorption lines, most cleanly with the new XRISM data. The derived wind properties match fairly well to that predicted by thermal-radiative driving by X-ray heating of the outer disc (e.g. Tomaru, this volume).
This mechanism predicts changes in the wind column and ionisation state as a function of the changing luminosity and spectral shape of the X-ray irradiation, broadly similar to what is seen in the data
(see e.g. \citep{Munoz-Darias2026}). 

The quasi-exponential lightcurves during the disk dominated decay from outburst peak do not show much additional faster variability superimposed e.g. such as would be associated with the thermal timescale. This is actually puzzling in the context of the SS73 disc model as these are unstable when radiation pressure dominates over gas pressure inside the disc. This is a consequence of the assumed viscosity prescription, where the heating scales with the total (gas plus radiation) pressure, rather than just the gas pressure or the geometric mean of gas and radiation pressure. This standard heating prescription gives rise to strong limit cycle variabilty for $\dot{m}=L/L_{\rm Edd}\ge 0.05$, where the system cycles between $0.01-2L_{\rm Edd}$ on the thermal/viscous timescale (the two are similar as the instability leads to an inner region with $H\sim R$) which is around 1000s (\cite{Szuszkiewicz2001}. This is not seen in general. The behaviour of GRS1915+104 and other extreme variability systems may be linked to this instability, but if so, it is triggered at much higher luminosity than predicted by the standard $\alpha$ disc models, around $L_{\rm Edd}$ rather than $0.05L_{\rm Edd}$ (see e.g. \citep{Grzedzielski2017}).  This is still an issue, as the  MHD simulations (see e.g. Scepi, this volume) still favour the standard heating prescription. 

Instead, there is a very obvious transition in the disc properties at $\dot{m}\sim 0.01$ on the slow decline from outburst. The spectrum goes from being dominated by the cool, thermal disc, peaking below $\sim 1~keV$, to being dominated by Comptonised emission from the corona, peaking at 100~keV. This most likely signals the transition to a very different type of accretion flow, one which is optically thin, hot, and geometrically thick \cite{Done2007}. These are the radiative inefficient accretion flows (RIAF), the best known of which are the Advection Dominated Accretion Flows (ADAF: \citep{Narayan1995}). This is another steady state solution of the accretion flow equations, whwre unlike the disc solution, the flow is not dense enough to thermalise. The ADAF solution assumes that the gravitational power goes mainly to the protons, heating them to their virial temperature.  The electrons gain energy by interacting with the protons, but they radiate it efficiently (as Comptonised emission), so the electron temperature is much lower than that of the protons. This two temperature flow is only possible at low mass accretion rates, as the density of the flow increases with mass accretion rate and eventually the electrons and ions interact sufficiently often to equilibriate. This happens roughly when the flow becomes optically thick, at a luminosity of $L=1.3\alpha^2 L_{\rm Edd}$ \cite{Narayan1995,Esin1997} which matches to the observed tranistion luminosity of $\sim 0.02L_{\rm Edd}$ \cite{vahdat-motlagh2019,maccarone2003} for $\alpha\sim 0.1$.

The ADAF flow is generally geometrically thick as the ion pressure is high, and the high ion temperature also sets the sound speed and viscous speed, so the mass accretion rate can vary very much faster in these flows than in an SS73 disc. Indeed, these hard spectra generically show fast (subsecond) variability, quite unlike the disc dominated data. This fast variability can also include a strong quasi periodic oscillations (QPOs), and the best current model for this is global Lense-Thirring (relativistic vertical) precession of the entire hot flow \cite{Ingram2009,Ingram2019}.

The large scale height of the hot flow means it is more able to carry large scale height magnetic field than the thin disc, giving an explanation for the
strong correlation of the radio jet with the hard coronal emission and not with the total masss accretion rate. 
In the systems showing the dramatic variabilty triggered by the DIM, the spectra which are dominated by the disc have no compact radio jet, whereas those in the Compton dominated state have radio emission which scales with the X-ray flux, as expected in a Blandford Konigal conical jet, giving the 'fundamental plane' of radio-X-ray behaviour (see below, \cite{Heinz2003,Merloni2003}). The observed correlation for the stellar mass black holes has a scatter of up to 1 order of magnitude, and there is now a suggestion that individual sources have higher or lower radio to X-ray ratios (see e.g. data from the compilation of \cite{Bahramian2022}). By analogy with the supermassive black holes, these are sometimes termed radio-loud and radio-quiet, respectively, but the range in radio to X-ray behaviour is far smaller than that seen in the supermassive black holes (see below). Instead, using the factor 10 difference in radio (and IR) to X-ray ratio as an upper limit means 
that beaming of the radio emission is not strong (Lorentz factor $\Gamma<3.5$) as otherwise these different inclination sources would 
show more difference in radio/IR jet emission relative to the (assumed) more isotropic coronal X-ray flux \cite{Saikia2019} (but see \cite{Lilje2025}).

This picture of a transition from an accretion flow dominated by a standard disc to a hot flow gives a fairly good overall picture of the evolution of the spectral and timing properties as a function of mass accretion rate as the source luminosity declines after the disk instability.  The jets tie into the X-ray emitting hot flow rather than the disc, and the winds are produced by thermal heating due to central X-ray (disc and coronal) irradiation of the outer disc and respond to the changing X-ray spectral shape. 

\section{Challenges to this picture}

There are three main issues which mean that this picture is still somewhat debated. 

Firstly, observationally, the hard spectra 
associated with a hot flow persist to high luminosities on the fast rise to the outburst peak. These bright hard state spectra are clearly above the standard ADAF limit, so their nature and origin are unclear (see below). Nontheless, it is difficult to maintain a hard spectrum in the vicinity of an optically thick disc as reprocessing of any X-ray illumination gives a source of seed photons for Compton cooling, which softens the spectra \cite{haardt1993,Poutanen2018,Ranjan-Datta2025}. 
These bright hard states are also the ones which show the strongest QPOs which again suggests that there is no inner thin disc present on the midplane, as this would otherwise prevent a vertical oscillation of the flow assuming that the QPO is set by Lense-Thirring precession. However, any sort of hot flow solution
to replace the inner disc out to some truncation radius is challenged by data which clearly show a broad base to the iron line produced by reflection of X-rays from the disc. This broad base remains quite constant in size and width across the transition from the bright hard state to a much softer, stronger disc state \citep{Kara2019}. These authors atttribute this to an optically thick inner disc remaining present, extending down to the last stable circular orbit of a high spin black hole 
even in the bright hard states. This apparently rules out inner disc truncation for these states. Their model to explain the spectral change instead has a persistant inner thin disc, but with a 
'lamppost' corona (compact/jetlike source on the spin axis above the black hole), which shrinks down as the spectrum softens. However, such a source geometry predicts
that the X-ray polarization should be dominated by reflection from the thin disc, predicting 
a polarisation angle which is parallel to the disc (and hence perpendicular to the jet) 
whereas data from IXPE now shows clearly that the polarisation is parallel to the jet. This is instead consistent with Compton scattering in a radially extended, optically thin, hot inner flow \citep{Krawczynski2022}.

The issue with reflection in the bright hard states is not just confined to a single object. A survey of all reflection fits to NuStar data (not affected by technical issues for observing bright sources) on stellar mass black holes gives generally high spins \citep{Draghis2024}. 
This is puzzling as the first results from direct detection of gravitational waves from merging black holes showed low spins (\citep{lvk2023}, see also \citep{Belczynski2024}).

The resolution of this issue may be connected to 
the ionisation state of the reflecting material. 
This is observed to be quite high, so the 
photosphere is hot, and there is substantial 
Compton up/downscattering of the line emission 
in this material. This gives a broad base to the 
line which is different only in detail to 
relativistic smearing (see e.g. 
\citep{Zdziarski2021}).
This is especially a source of uncertainty in 
the stellar mass systems as the disc itself is 
X-ray hot. The vertical and radial temperature/ionisation structure of the reflecting material is set not just by illumination from above as assumed in the models fit so far, but should also include the intrinsic disc flux from below \citep{Ding2024}. More unexpectedly, these authors also additionally claim significant (factor of 7 increase in the intensity!) differences in the line profiles of strongly ionised iron when the reflection models are redone using new atomic databases. If this is real issue, rather than incomplete implementation of the new atomic data then it clearly has implications for all ionised reflection results, in AGN as well as in binaries. Understanding the extent and applicability of these changes in atomic data is a top priority to assess the accuracy of inner radii determined by reflection modelling from highly ionised material but there is now a new public reflection code using the new atomic data so there should be progress on this issue in the near future \citep{huang2025}.

Secondly, theoretically, there are no well established models which can give a detailed picture of {\it any} of the bright accretion flow states. The standard disc models (where heating is proportional to total pressure) should show radiation pressure instability induced limit cycles above $\dot{m}=0.05$ \citep{Szuszkiewicz2001}, but the data do not, and instead support a modified heating prescription (see e.g. \citep{Grzedzielski2017}). The luminosity of the bright disc state also fades on timescales of $\sim 30$ days, which requires efficient angular momentum transport, with an effective $\alpha>0.1-0.2$ even when including thermal winds \citep{Avakyan2024}, while the MRI simulations with no net vertical flux give values an order of magnitude lower (see Scepi, this volume and \citep{Salvesen2016}). 

Conversely, in addition to the issues of geometry discussed above, an 
ADAF should collapse at $L\sim 0.02L_{\rm Edd}$. Yet the data show bright hard states up to $\sim 0.2L_{\rm Edd}$ and even above. These are seen mainly on the fast rise to outburst in the DIM, so are most likely out of steady state, yet the hydrodynamic timescales for the flow to collapse are much shorter. Magnetic pressure seems to be the only solution, and would also probably give the larger $\alpha$ required by the observations as the two seem to be connected \citep{Salvesen2016}, but the details of the resulting magnetic configuration are still very unclear. Strong ordered fields are difficult to reconcile with the observed X-ray polarization, as they predict Faraday rotation with depolarises the emission, whereas the data show a polariation which is already quite high compared to expectations \citep{Barnier2024}. 

Thirdly, again theoretically, there is currently no consensus on how the transition from disc to flow can take place to give a truncated disc-hot inner flow geometry. The early models stressed hydrodynamics, with disc evaporation as the mechanism \citep{Meyer1994,Rozanska2000,
Mayer2007},
while later work has used magnetic pressure from large scale fields to truncate the disc 
(Magnetically Arrested Disc: MAD)
in GR MHD simulations \citep{Liska2022}. There are however several issues with MAD flows, firstly, they have strong ordered magnetic fields threading the hot flow region, which probably produces too much Faraday rotation to see any intrinsic X-ray polarisation through Compton scattering, as discussed above \citep{Barnier2024}. Secondly, to produce a MAD seems to require that the flow starts with net magnetic flux, otherwise it instead forms a SANE (standard and normal evolution) hot flow threaded with turbulent field from the small scale dynamo rather than large scale ordered fields \citep{Begelman2022,Liska2022,Tchekhovskoy2011}. Net flux strongly changes the jet properties, in a way that does not seem to match well with the rather tight radio-X-ray correlation of the stellar mass black holes, and its extrapolation on the Fundamental Plane to AGN (see below). 

There have been some more recent attempts to calculate a radial transition from a thin disc to an ADAF using more modern hydrodynamic calculations \citep{Nemmen2024,Hogg2017,Wu2016,Das2013}, and it may be that this approach gives more understanding of the mechanisms involved. 

\section{Supermassive black holes}

This picture of the accretion flow built from observations of the DIM in stellar mass black holes is not easy to scale up to the supermassive black holes where the accretion powers the Active Galactic Nuclei (AGN) and Quasars. 
The SS73 disc model has peak temperature scaling as $T\propto (\dot{m}/M)^{1/4}$, where $\dot{m}=L/L_{\rm Edd}$, so a $10^9M_\odot$ black hole at $\dot{m}\sim 0.5$ has disc temperature which is $100\times$ lower than for a similarly accreting $10M_\odot$ black hole. This predicts a peak in $\nu f_\nu$ at $35$~eV rather than $3.5$~keV as seen in the stellar mass black holes. This is in the unobservable far UV part of the spectrum, so the disc can only be seen directly at larger radii, rather than at the peak itself. 

Since the inner disc temperature is only just above Hydrogen ionisation, it is almost inevitable that the outer disc will drop below the Hydrogen instability point 
\citep{burderi1898}.
In the stellar mass accreting objects, this condition is sufficient to trigger the DIM. 
However, the DIM may not operate in the same way in supermassive black hole discs, as there may not be a switch in the $\alpha$ effective viscosity, from $\alpha_{cold}\sim 0.02$ to $\alpha_{hot}\sim 0.2$. Plausibly, the drop in $\alpha$ at low temperatures in the stellar mass systems is due to the drop in charge carriers to couple the magnetic field to the gas
below Hydrogen ionisation temperatures. 
This drop in conductivity means that resistivity (a non-ideal MHD effect) becomes important, and can completely suppress the MRI dynamo 
\citep{Menou2001,Scepi2018}. 
Instead, in the lower densities of AGN discs, ambipolar diffusion be more important than resistive diffusion, keeping the neutrals well coupled to the field by collisions with ions \citep{Menou2001}. 

Other issues are that the 
lower SS73 disc temperature at a given $R/R_{\rm g}$ and $L/L_{\rm Edd}$ means gas pressure inside the disc is lower, so radiation pressure dominates at all reasonable $L/L_{\rm Edd}$ for bright AGN. This should lead to the disc being radiation pressure unstable, and to limit cycle behaviour - but since we do not see this in the stellar mass black holes, it is not at all clear what might happen in the supermassive black holes in this regime
(see e.g. the discussion in \citep{Blaes2025}). 

The variability we do see in the disc dominated stellar mass black holes is on the viscous timescale, but this should be extremely long for AGN. Even the transition timescale, where the inner disc is replaced by the hot flow, takes of order 0.5~days in the stellar mass systems 
\citep{Dunn2010}, so should take a million times longer i.e. 3000 years even for a low mass AGN at $10^7M_\odot$ (e.g. \citep{Noda2018}). Thus we do not expect to see the equivalent of state transitions in individual AGN. Yet we do. 

To date, this is most clearly shown in Mkn~1018, where SWIFT snapshot observations show that the broad band spectral energy distribution (SED) goes from peaking in the UV to being dominated by the hard X-rays \citep{Noda2018}. This looks very like the evaporation of the disc into a hot flow seen in the transition in the stallar mass black holes. The drop in the UV also means that the broad line region flux drops, giving the term 'changing look' AGN as a single object goes from showing a strong UV continuum and broad H$\beta$ to showing no (or much less) UV and no (or much less) broad line. Unified AGN models explained AGN with no UV/broad lines as obscured, yet this change is not due to changing obscuration as the X-rays are clearly unobscured, and in many objects the IR (which is reprocessed UV from the torus)
clearly also drops \citep{Lyu2021}, again showing this is an intrinsic change in the source luminosity. The best estimates for where this transition happens is $\sim 0.01L_{\rm Edd}$, as seen in the stellar mass systems on their slow decline (see e.g. \citep{Jana2026} and references therein). 

The transition can also be seen in the demographics of AGN picked out in large surveys e.g. SDSS. There are hundreds of thousands of AGN at $L\sim 0.1-0.3L_{\rm Edd}$ across a wide range of masses $10^{7-9}M_\odot$, but the numbers drop abruptly below $0.01L_{\rm Edd}$, as expected if the AGN lose their strong blue disc continuum and broad lines below this point \citep{Mitchell2023}. However, this is more complex to interpret as it becomes more difficult to separate out the AGN from the host galaxy. Dust reddening could also reduce the observed emission from the 
AGN, and it is difficult to separate this from an intrinsic change in the source spectrum. These should actually be correlated, as the AGN illumination of the dusty torus should drive a thermal-radiative wind. Above the transition, the bright central UV will give additional radiative acceleration of the wind, so it is blown away quite efficiently. Just below the transition, the hard X-ray illumination gives only a thermal wind, so the entrained dust can reach a larger scale height (see e.g. \citep{Hoenig2017}). Thus AGN at and below the transition, with no bright UV emission, should be more likely to be obscured/reddened than bright AGN. This is seen in e.g. hard X-ray surveys of AGN \citep{Ricci2022}.

The eROSITA soft X-ray survey now gives an unbiased way to select unobscured ($N_H\le 10^{22}$~cm$^{-2}$) AGN. Its eFEDs field in particular is the deepest, wide angle soft X-ray survey, with copious ancilary data, most notably HyperSuprimeCam (HSC) imaging in multiple optical filters. Unobscured AGN are then easily identified 
as soft X-ray point sources in the centre of an optically imaged host galaxy. HSC covers the whole eFEDS area, and has excellent imaging at around $0.4"$ resolution. This enables detailed host galaxy modelling, as the images can separate out the central point source from the extended galaxy, giving the AGN SED without host galaxy contamination. The host galaxy images can also be modelled in detail to get the uncontaminated galaxy luminosity, and hence an estimate for black hole mass which can be used consistently across the entire sample, unlike the single epoch BLR spectroscopic masses which can only be used where there is a BLR.  Selecting only a single host galaxy stellar mass bin around $10^{11}M_\odot$ (where the host galaxy subtraction is most reliable)
means selection on a single black hole mass bin $10^{8-8.5}M_\odot$, and stacking 
on the HSC AGN optical luminosity for these gives a sequence of optical-X-ray SEDs as a function of $L/L_{\rm Edd}$. These very clearly show the spectral transition. Below $0.01L_{\rm Edd}$ the SED is dominated by hard X-rays, with very little UV, while above  $0.01L_{\rm Edd}$ there is a growing UV component
\citep{Hagen2024,Kang2025}.
The corresponding SDSS spectroscopic data (now covering the lower $L/L_{\rm Edd}$ sources as these objects were targetted as eROSITA sources) shows clearly that the BLR appears alongside the UV continuum component \citep{Kang2025}.

This sample of AGN (constant black hole mass, changing Eddington ratio across the transition) also allows us to explore what happens to the radio jet. In the DIM triggered stellar mass black holes, the hot flow persists to luminosities far above the expected transition at $0.01L_{\rm Edd}$, and is generally a switch to a disc dominated state. The compact steady radio jet emission then collapses. However, in the non-DIM black holes (i.e. Cyg X-1) the switch is less dramatic. There is always a fairly substantial X-ray tail even in the softest state, and the radio emission from the compact jet clearly tracks these high energy coronal X-rays rather than following the total emission which is dominated by the optically thick component \citep{Zdziarski2020}.

We use the new VLASS all sky radio survey data to get the radio emission for this eFEDS-HSC-Galex sample of AGN SEDs. Very few of these sources are individually detected, but stacking the radio images gives detections for 3 SED bins, one below the transition, one just above, and one at higher $L/L_{\rm Edd}\sim 0.1$. These show very similar radio luminosities, despite more than an order of magnitude change in total AGN luminosity, but, like Cyg X-1, the radio flux follows the X-ray coronal flux i.e. ties into the optically thin, hot flow component not to the optically thick disc emission \citep{Kang2025b}. 

\section{Fundamental plane: unifying the disk and its jet across the mass scale}

There is a fixed relationship predicted between the radio flux from a Blandford-Konigl (conical, self similar) jet and the X-ray flux expected from an ADAF if the jet power, $P_{\rm jet}$, is directly proportional to the accretion power, $P_{\rm acc}$. The vertically extended radio jet should emit a range of synchrotron self absorbed spectra, which sum together giving a predicted monochromatic radio flux of $L_R\propto (\dot{m}M)^{17/12}$. Assuming that the accretion X-ray emission is from an ADAF then this has $L_x\propto \dot{m}^2M$ as the ADAF radiative efficiency increases with density ($\propto \dot{m}$) as this increases the rate at which the electrons interact with the protons. Substituting for $\dot{m}$ gives the fundamental plane relation 
$L_R\propto [(L_x/M)^{1/2} M]^{17/12} \propto (L_x M)^{17/24}$ \citep{Merloni2003}.
This is the simplest possible, physically motivated scaling between $L_R$ (monochromatic) and $L_X$ (total coronal luminosity) across different black hole masses and mass accretion rate.

\begin{figure}
    \centering
	\includegraphics[width=1.05\columnwidth]{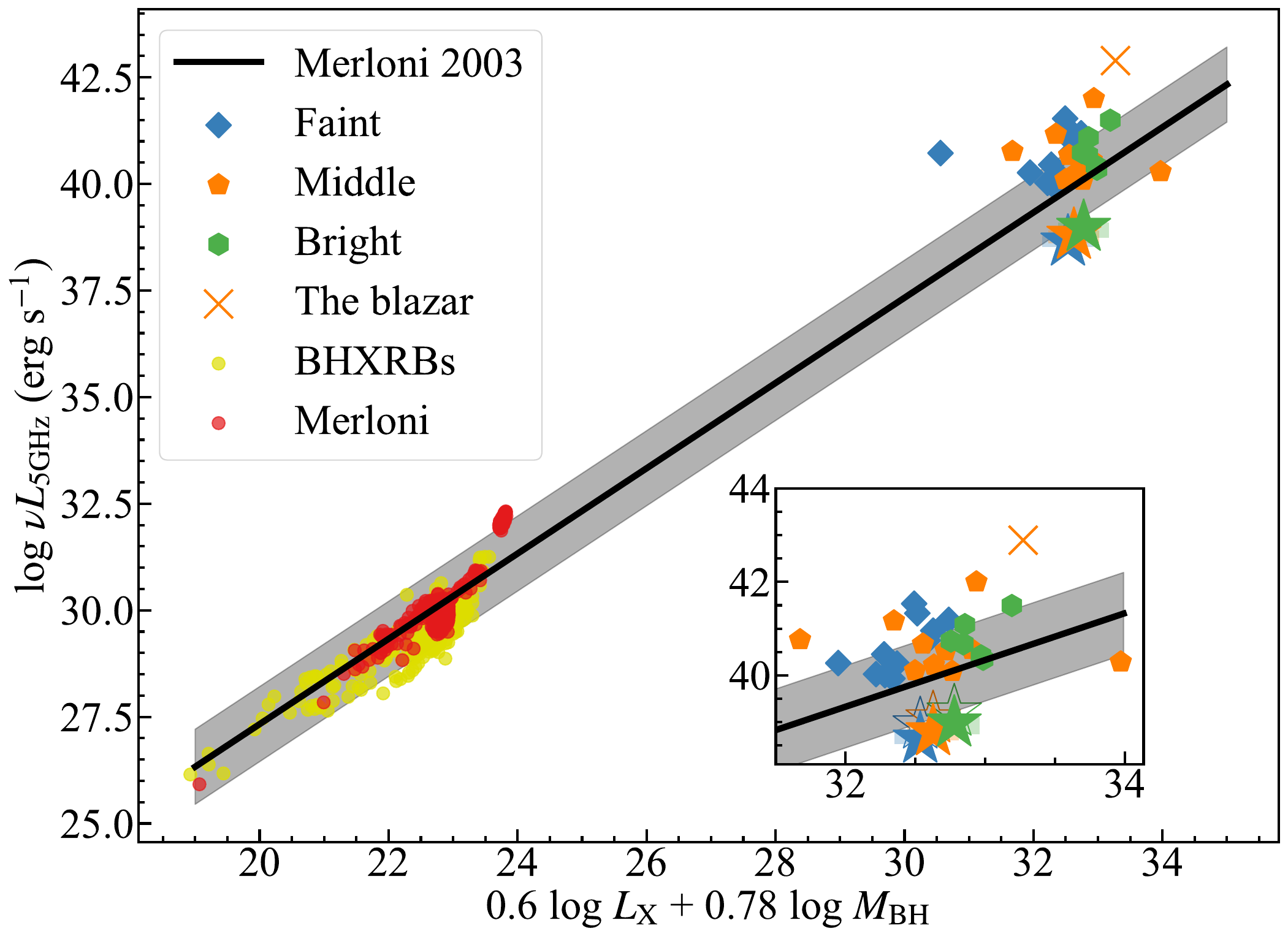}
  
    \caption{Fundamental plane from stellar mass black holes\citep{Bahramian2022} with "radio loud" in red (almost all of the original sample in \citep{Merloni2003}), "radio quiet" in yellow. These are actually within an order of magnitude of each other. Instead, the stars show the stacked non-detected in radio (VLASS) AGN sample in three mass accretion rate bins (blue: $L/L_{\rm Edd}\sim 0.01$, orange: $L/L_{\rm Edd}\sim 0.02$, green: $L/L_{\rm Edd}\sim 0.07$) are all slightly below the extrapolated relation from all the stellar mass black holes. This contrasts with the few individually detected VLASS radio sources (faint: diamonds, middle: hexagons, bright: circles). 
    The inset shows a zoom-in of the AGN. These have a factor 100-1000$\times$ larger radio flux {\em for the same range of mass and mass accretion rate and large scale environment as the stacked non-deteced AGN}. Clearly there is another parameter in addition to mass and mass accretion rate  which controls the jet power \citep{Kang2025b}.
    }
    \label{fig:FP}
\end{figure}

The early compilations of accretion flows with both X-ray and radio detections were indeed consistent with lying on a single $M-L_X-L_R$ plane \citep{Merloni2003}, across both AGN and the stellar mass black holes. With better data, the stellar mass black holes are now often split into 'radio loud' and 'radio quiet' subgroups, but these are less than a factor 10 different, which is not a large distinction on the fundamental plane \citep{Bahramian2022}.
Our stacked AGN sample lie slightly below the predicted relation, by an amount which is slightly bigger than the 'radio quiet' stellar mass back holes \citep{Kang2025b}.

However, the very 
few individually detected 'radio loud' AGN in our sample lie orders of magnitude above, for black holes of the same mass and mass accretion rate, and in the same (on average) large scale environments. This is important when comparing very massive AGN, as the most massive black holes live in the most massive galaxies which are often central cluster galaxies, surrounded by a high pressure hot gas halo. This gives a much stronger working surface for the jet, so a larger fraction of the kinetic power can be converted to radio emission. This would be expected to affect the larger scale emission from the lobe rather than the core, but issues of spatial resolution then arise. Instead, 
our sample, by selecting on host galaxy stellar mass, avoids dispersion from strongly clustered environments. Instead, our dispersion is real: AGN with the same black hole mass and spanning the same range of black hole mass accretion rates, give very different radio jet powers. 

The obvious question is why this occurs. 
In the simplest possible models, this could be due to black hole spin. The X-ray-radio correlation on the fundamental plane for the majority of (radio quiet) AGN at $10^{8-8.5}M_\odot$ and the stellar mass black holes then would indicate that these sources all have similarly (lowish) spins, while the few individally detected objects have much higher spin. The theoretical power from Blandford-Znajek (spin powered) jets can vary by a factor of more than 4 orders of magnitude \citep{Unal2020}, giving a shift in proportionality factor between $P_{\rm jet}$ and $P_{\rm acc}$ which is sufficient to match the observed spread \citep{Unal2020}. This is in principle testable, as spin is encoded on the spacetime along with mass, making it observationally accessible via its effect on the last stable circular orbit in Einsteins gravity, setting the inner radius of an (untruncated) accretion disk. However, as discussed in Section 2, there is considerable controversy over how to interpret reflection results giving the inner radius of discs even in stellar mass black hole binaries with excellent data. Reflection fits to AGN currently also give high spins on average, 
\citep{Reynolds2021}, though these objects are generally on the 'radio quiet' side of the Fundamental Plane. If the ionised reflection models are appropriate (but see section 2) then this rules out the simplest spin-jet paradigm and another parameter is required. 

Jet power in the GR-MHD simulations also strongly depends on the net magnetic flux \citep{Tchekhovskoy2011,Sikora2013}, so the rare 'radio loud' quasars could be triggered instead by a more complex combination of high spin and high magnetic flux.

\section{Summary}

Black hole accretion flows should depend on just mass and spin, together with mass accretion rate. Inclination can determine some aspects of how the flow is seen, and there can also be some non-steady state and/or environmental effects, but fundamentally there are only three parameters. The effects of mass and (mass scaled) accretion rate are plainly apparent in the data, especially in the existance of the spectral transition to an X-ray hot flow below $\dot{m}\sim 0.01$. 
This transition does not in itself determine the difference between radio-loud and radio quiet quasars, and it is possible that this instead due to a small fraction of AGN having  
high black hole spin, producing much stronger jet powers than standard low/moderate spin jets. It seems useful to explore this simplest version of a model which can explain the observed zoo of AGN properties before going to the currently untestable solution where jet power depends on the history of accretion of magnetic flux.

\section*{Acknowledgments}
I thank my many collaborators who contributed to my understanding of accretion, e*specially Scott Hagen who worked on the original eROSITA-HSC sample, and JiaLia Kang who extended it into the UV and radio. I acknowledge support from STFC via grant ST/T000244/1, and 
thank the Fujihara foundation for financial support which enabled me to attend the workshop. 

\bibliographystyle{JHEP}
\bibliography{skeleton}

\end{document}